# Nanoparticles designed from low pressure plasmas as model particles for astrophysical dust

I. Stefanović[1,2], E. Kovačević[1], J. Berndt[1], Y.J. Pendleton[3], and J. Winter[1]

1. Institute of Experimental Physics II, Ruhr-University Bochum, 44780 Bochum, Germany
2. Also at Institute of Physics, POB 57, Belgrade, Serbia and Montenegro
3. NASA Ames Research Center, Mail Stop 245-3, Moffett Field, CA 94035, USA

Dust particles are ubiquitous in space, including e.g. diffuse and dense interstellar media (ISM), nova ejecta, the outflow of red giant stars and accretion disks, proto-planetary nebula etc. The identification of solid hydrogenated carbonaceous material in the diffuse ISM has been confirmed by infrared (IR) observations and has been studied in laboratory [1].

An important tool for the interpretation of astronomical IR data and the understanding of the observed phenomena is the availability of IR spectra gained from laboratory investigations on analogue materials. Plasma polymerization of hydrocarbons (e.g. argon diluted acetylene) is an efficient source for dust formation in laboratory. The resulting IR spectra reveal the aliphatic CH stretch at 3.4 μm as well as traces of aromatic compounds and a weak OH and carbonyl presence [2]. A comparison of the 3.4 μm band profiles of the nanoparticles bands wit those that originate from circumstellar material [2, 3, 4] reveals striking similarities. Furthermore, these experimental IR fingerprints are also similar to the 2 – 10 μm spectra of dust in the diffuse ISM, e.g. IR spectra of dust as seen towards luminous background sources such as the Galactic Center [1, 2, 3, 5]. Our polymerization method enables us to vary, in a simple and controlled way, the physical and chemical parameters important for particle and chemical formation. In the case of the particles grown with nitrogen as a carrier gas, the spectra show strong dominance of nitrogen bonds over pure hydrocarbon bonds, producing a band at 4.62 μm, which has been identified in dense cold dust [6, 7]. Plasma induced growth of dust also allows investigation of different in-situ measurements on the collective effects of the dust (e.g. charging processes) and on the growth dynamics.